\documentclass{elsart}

\begin{document}
\begin{frontmatter}
\centerline{Theoretical Computer Science, 328/1-2(2004), 151-160} 
  \title{Reducing the time complexity of testing for local threshold
 testability}
 \author{A.N. Trahtman}
\date{}
\ead{trakht@macs.biu.ac.il}
 \address{Bar-Ilan University, Dep. of Math., 52900, Ramat Gan, Israel}
\begin{abstract}
 A locally threshold testable language $L$ is a language with
 the property that  for some nonnegative integers $k$ and $l$
 and for some word $u$ from $L$, a word $v$ belongs to $L$ iff
\\
 (1) the prefixes [suffixes] of length $k-1$ of words $u$
 and $v$ coincide,
\\
 (2) the numbers of occurrences of every factor
  of length $k$ in both words $u$ and $v$ are either the same or
 greater than $l-1$.
\\
 A deterministic finite automaton is called locally threshold testable
if the automaton accepts a locally threshold testable language for some
$l$ and $k$.\\
  New necessary and sufficient conditions
for a deterministic finite automaton
to be locally threshold testable are found.
On the basis of these conditions,
 we modify the algorithm to verify local
threshold testability of the automaton and to reduce the time
complexity of the algorithm.
The algorithm is implemented as a part of the
$C/C ^{++}$ package TESTAS.
\texttt{http://www.cs.biu.ac.il/$\sim$trakht/Testas.html}.
\end{abstract}
\end{frontmatter}
 \section{Introduction}
  The locally threshold testable languages introduced
by Beauquier and
 Pin \cite {BP} now have various applications \cite {REG},  \cite {VCG},
 \cite {W}. In particular,
stochastic locally threshold testable languages, also known as
 $\it n-grams$, are used
 in pattern recognition and in speech recognition,
both in acoustic-phonetics decoding and in language modelling
\cite {VCG}.
These languages generalize the concept of
local testability  \cite {BS}, \cite {MP},
which can be considered as a special case of
local $l$-threshold testability for $l=1$. \\
 Necessary and sufficient conditions of local testability
 \cite {K91} form a basis of polynomial-time algorithms for
the local testability problem \cite{K91}, \cite {TC}.
The algorithms were implemented \cite {C}, \cite {C2}, \cite{TC}.\\
Necessary and sufficient conditions of local threshold testability
for deterministic finite automata (DFA) found in \cite {BP} are
based on a syntactic characterization of locally threshold testable
languages \cite {TW}.
A polynomial-time algorithm of order $O(|\Gamma|^5|\Sigma|)$ for the
local threshold testability problem based on some other kind
of necessary and sufficient conditions was described in \cite{TC}
and implemented. We modify the last necessary and sufficient conditions
and reduce in that way the order of the algorithm for
local threshold testability to $O(|\Gamma|^4|\Sigma|)$.
The algorithm was successfully implemented.
\\
 \section{Notation and definitions}
 Let $\Sigma^+$ [$\Sigma^*$] denote the free semigroup [monoid]
over an alphabet $\Sigma$.
\\
 If $w \in \Sigma^+$, let $|w|$ denote the length of $w$.
Let $k$ be a positive integer. Let $i_k(w)$ $[t_k(w)]$ denote
the prefix
[suffix] of $w$ of length $k$ or $w$ if $|w| < k$. Let
$F_{k,j}(w)$ denote
 the set of factors of $w$ of length $k$ with at least $j$
occurrences.
   A language $L$
 is called {\it l-threshold k-testable} if there is an
alphabet $\Sigma$  such that for all
  $u$, $v \in \Sigma^+$, if $i_{k-1}(u)=i_{k-1}(v)$, $t_{k-
1}(u)=t_{k-1}(v)$
 and $F_{k,j}(u)=F_{k,j}(v)$ for all $j \le l$, then either
both $u$ and $v$
 are in $L$ or neither is in $L$. \\
   An automaton is {\it $l$-threshold $k$-testable} if the
automaton accepts a
 $l$-threshold $k$-testable language.
   A language $L$ [an automaton] is
{\it locally threshold} {\it testable} if it is
 $l$-threshold $k$-testable for
some $k$ and $l$.\\
    \hfill Let us now consider the transition graph of a DFA.\\
The action of a word $v \in \Sigma^*$ on a state $\bf q$ is denoted by ${\bf q}v$.
Thus ${\bf q}v$ is the state reached by the unique path of label $v$
starting at $\bf q$.
\\
A state $\bf p$ is a $\it cycle$ $\it state$
if, for some $e \in \Sigma^+$, ${\bf p}e=\bf p$.
\\
A maximal strongly connected component of a directed graph will be
denoted for brevity by $\it SCC$.
  \\
We shall write $\bf p \succeq \bf q$ if $\bf q$ is reachable from $\bf p$
(that is, if ${\bf p}v=\bf q$ for some word $v \in \Sigma^*$)
and $\bf p \sim q$ if $\bf p \succeq q$ \& $\bf q \succeq p$
(that is, if $\bf p$ and $\bf q$ are in the same $SCC$).
\\
The number of vertices of a graph $\Gamma$ is denoted by $|\Gamma|$.\\
 An oriented labelled graph is {\it complete} if any of its vertex
 has outgoing edge with any label from the alphabet of
labels. A non-complete graph can be completed by adding a sink
state and then adding lacking edges from corresponding
vertices to the sink state.\\
The direct product $\Gamma^k$ of $k$ copies of a directed labelled
graph $\Gamma$ over an alphabet $\Sigma$ consists of vertices $({\bf
p}_1, ..., {\bf p}_k)$ and edges (${\bf p}_1, ..., {\bf p}_k) \to
({\bf p}_1\sigma, ..., {\bf p}_k\sigma)$ labelled by $\sigma$.
Here ${\bf p}_i$ are vertices from $\Gamma$, $\sigma \in \Sigma$.
 \section{The necessary and sufficient conditions of local
threshold testability}
   Let us formulate the result of Beauquier and Pin \cite
{BP} in the
   following form:
   \begin{thm} $\label{1}$ \cite {BP}
    A language $L$ is locally threshold
   testable if and only if the syntactic semigroup $S$ of $L$
is aperiodic
and for any two idempotents $e$, $f$ and elements $a$, $u$,
$b$ of $S$, we have
    \begin{equation}
                    eafuebf=ebfueaf.
                               \label {e1}
   \end{equation}
  \end{thm}
We now consider a fixed locally threshold testable DFA
 with state transition graph $\Gamma$ and transition semigroup $S$.
   \begin{lem} $\label{2}$ \cite {K91} \cite {TC}
  Let ($\bf p, q$) be a cycle state of $\Gamma^2$.
If ${\bf p} \sim \bf q$, then ${\bf p} = \bf q$.
   \end{lem}
\begin{lem} $\label{4}$
  Let (${\bf q, t}_1$) and (${\bf q, t}_2$) be cycle states
   of $\Gamma^2$.
     If $({\bf q, t}_1) \succeq ({\bf q, t}_2)$
 and ${\bf q} \succeq {\bf t}_1$ then ${\bf t}_1 \sim {\bf t}_2$.
   \end{lem}
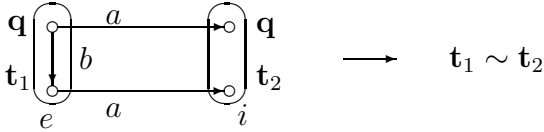
\begin{picture}(170,44)
\put(35,18){\circle{4}}
      \put(17,20){${\bf t}_1$}

\put(55,8){$a$}
\put(55,43){$a$}
\put(45,26){$b$}

\put(37,42){\vector(1,0){63}}
\put(37,18){\vector(1,0){63}}

\put(102,42){\circle{4}}
 \put(112,40){$\bf q$}
 \put(35,40){\vector(0,-1){20}}
\put(35,42){\circle{4}}
 \put(18,41){$\bf q$}
\put(102,18){\circle{4}}
\put(112,20){${\bf t}_2$}

\put(35,32){\oval(14,38)}

 \put(30,4){$e$}

\put(101,32){\oval(14,38)}
\put(105,5){$i$}

 \put(145,30){\vector(1,0){20}}
  \put(185,27){${\bf t}_1 \sim {\bf  t}_2$}
 \end{picture}\\
Proof.
One has (${\bf q, t}_1)e = ({\bf q, t}_1$),
(${\bf q, t}_2)i= ({\bf q, t}_2$),
(${\bf q, t}_1)a = ({\bf q, t}_2$),
${\bf q}b = {\bf t}_1$ for some idempotents $e$, $i$
and elements $a$, $b$ from $S$.
The substitution $ai$ in place of $a$ and $e$ in place of $f$ and $u$
 in (\ref{e1}) implies $eaiebe=ebeaie$.
Therefore ${\bf t}_2e = {\bf t}_2ie = {\bf t}_1eaie =
{\bf q}ebeaie ={\bf q}eaiebe$.
Thus ${\bf t}_2e = {\bf q}eaiebe={\bf q}ebe = {\bf t}_1e ={\bf t}_1$.
So ${\bf t}_2 \succeq {\bf  t}_1$.
We have ${\bf t}_1a = {\bf t}_2$, whence ${\bf t}_1 \sim {\bf t}_2$.
\begin{lem} $\label{5}$
Let  ${\bf p, q, t, r, s}$ be states
such that $({\bf p, s})$ and (${\bf r,  t}$)
are cycle states of $\Gamma^2$. If $({\bf p,s}) \succeq ({\bf q, t})$
  and  ${\bf p} \succeq {\bf r} \succeq {\bf s}$,
then ${\bf q} \succeq {\bf t}$.
 \end{lem}
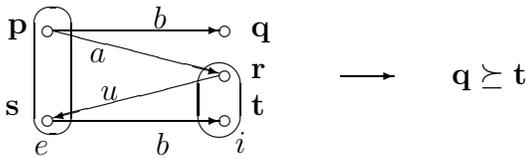
\begin{picture}(170,54)
\put(35,18){\circle{4}}
      \put(19,20){${\bf s}$}

 \put(112,34){$\bf r$}
\put(76,5){$b$}
\put(75,53){$b$}
 \put(100,35){\vector(-4,-1){63}}
\put(55,26){$u$}
\put(51,40){$a$}

\put(102,35){\circle{4}}
\put(37,52){\vector(1,0){63}}
\put(37,18){\vector(1,0){63}}

\put(102,52){\circle{4}}
 \put(112,50){$\bf q$}
 \put(37,52){\vector(4,-1){63}}
\put(35,52){\circle{4}}
 \put(20,51){$\bf p$}
\put(102,18){\circle{4}}
\put(112,20){$\bf t$}

\put(37,37){\oval(14,48)}

 \put(30,5){$e$}

\put(100,26){\oval(16,28)}
\put(106,6){$i$}

 \put(146,35){\vector(1,0){20}}
  \put(188,32){${\bf q} \succeq {\bf  t}$}
 \end{picture}
\\
Proof. One has $({\bf p, s})e=({\bf p, s})$ and
   (${\bf r},{\bf t})i=({\bf r},{\bf t})$
  for some idempotents $e, i \in S$.
Furthermore, $({\bf p,s})b=({\bf q, t})$,
  ${\bf p}a={\bf r}$ and ${\bf r}u={\bf s}$
  for some elements $a$, $u$, $b \in S$.
In view of (\ref{e1}), ${\bf t}={\bf p}eaiuebi = {\bf p}ebiueai$.
 Thus ${\bf t} = {\bf p}ebiueai = {\bf q}iuebi$,
whence ${\bf q} \succeq {\bf t}$.
\begin{lem} $\label{6}$
  Let ($\bf q, r$), (${\bf p, s}$), (${\bf q, t}_1$) and (${\bf q, t}_2$)
be cycle states of the graph $\Gamma^2$ such that
 $({\bf p, s}) \succeq ({\bf q, t}_i)$,
${\bf q} \succeq {\bf t}_i$ for $i=1,2$
    and ${\bf p} \succeq {\bf r} \succeq {\bf s}$.
   Then ${\bf t}_1 \sim {\bf t}_2$.
 \end{lem}
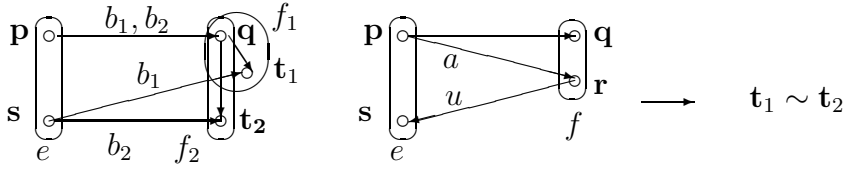
\begin{picture}(150,70)
\put(25,30){\circle{4}}
  \put(9,30){${\bf s}$}
 \put(97,27){$\bf t_2$}
\put(109,66){$f_1$}

 \put(27,30){\vector(1,0){63}}
\put(90,30){\circle{4}}
  \put(47,17){$b_2$}

\put(90,62){\circle{4}}
 \put(96,61){$\bf q$}
 \put(28,62){\vector(1,0){61}}
\put(90,60){\vector(0,-1){29}}
\put(93,62){\vector(2,-3){9}}
\put(25,62){\circle{4}}
\put(10,61){$\bf p$}
  \put(46,65){$b_1, b_2$}

 \put(25,30){\vector(4,1){73}}
\put(58,44){$b_1$}

\put(100,48){\circle{4}}
\put(110,46){${\bf t}_1$}
\put(96,56){\oval(24,30)}

\put(25,46){\oval(10,46)}
\put(90,46){\oval(10,46)}
 \put(20,15){$e$}
 \put(72,16){$f_2$}

\end{picture}
\begin{picture}(180,70)

\put(5,30){\circle{4}}
      \put(-12,30){${\bf s}$}

 \put(77,40){$\bf r$}

 \put(70,45){\vector(-4,-1){63}}
\put(70,45){\circle{4}}
\put(70,62){\circle{4}}
 \put(77,60){$\bf q$}
 \put(7,62){\vector(1,0){63}}
\put(7,62){\vector(4,-1){63}}
\put(5,62){\circle{4}}
\put(-10,61){$\bf p$}

\put(21,36){$u$}
\put(20,50){$a$}

\put(5,46){\oval(10,46)}
\put(69,53){\oval(10,30)}
 \put(0,15){$e$}
 \put(66,25){$f$}

\put(95,37){\vector(1,0){20}}
  \put(136,35){${\bf t}_1 \sim {\bf t}_2$}
 \end{picture}\\
Proof. One has
   $({\bf p, s})e=({\bf p, s})$,
     $({\bf q, r})f=({\bf q, r})$,
   (${\bf q},{\bf t}_1)f_1=({\bf q},{\bf t}_1)$,
   (${\bf q},{\bf t}_2)f_2=({\bf q},{\bf t}_2)$
  for some idempotents $e$, $f$, $f_1$, $f_2 \in S$,
 Furthermore, $({\bf p,s})b_1=({\bf q, t}_1)$,
   $({\bf p,s})b_2=({\bf q, t}_2)$,
   ${\bf p}a={\bf r}$ and ${\bf r}u={\bf s}$
  for some elements $a$, $u$, $b_1$, $b_2 \in S$.\\
\hfill   Let us consider the state ${\bf t}_if$ ($i=1,2$) where the
idempotent $f$ is right unit for $({\bf q, r})$.
The states $({\bf q, t}_i)$ and $({\bf q, t}_if)$
are cycle states,
 ${\bf q} \succeq {\bf t}_i$,
$({\bf q,t}_i) \succeq ({\bf q, t}_if)$,
whence by Lemma \ref{4}, ${\bf t}_if \sim {\bf t}_i$ for any such $f$.
So ${\bf t}_1f \sim {\bf t}_1$ and ${\bf t}_2f \sim {\bf t}_2$.
\\
 The equality of local
threshold testability (\ref{e1}) implies ${\bf t}_1f ={\bf s}b_1f ={\bf
r}ueb_1f = {\bf p}eafueb_1f= {\bf p}eb_1fueaf$.  Furthermore, ${\bf
p}b_1 = {\bf p}b_2= {\bf q}$, whence ${\bf t}_1f = {\bf p}eb_1fueaf
= {\bf p}eb_2fueaf = {\bf p}eafueb_2f= {\bf t}_2f$. So  ${\bf t}_2f =
{\bf t}_1f$. We have ${\bf t}_1f \sim {\bf t}_1$ and ${\bf t}_2f \sim {\bf t}_2$,
whence ${\bf t}_2 \sim {\bf t}_1$.\qed

 \hfill If ${\bf p, q, s}$ are states of  $\Gamma$ and there exists
some state ${\bf r}$
such that $({\bf q, r})$ and $({\bf p, s})$
are cycle states of $\Gamma^2$,
${\bf p} \succeq {\bf q}$,
 and ${\bf p} \succeq {\bf r} \succeq {\bf s}$, then the
non-empty set\\
 \centerline{
$ T= \{ t$ $| ({\bf p,s}) \succeq ({\bf q, t})$, ${\bf q} \succeq {\bf t}$
and (${\bf q,  t}$) is a cycle state$\}$
  }\\
by the previous Lemma is a subset of some $SCC$ from transition graph $\Gamma$
of locally threshold testable automaton.
This $SCC$ will be denoted by $SCC({\bf p, q, s})$. In the case
that $\bf r$ does not exist or $T$ is empty,
let $SCC({\bf p, q, s})$ be empty.
\\
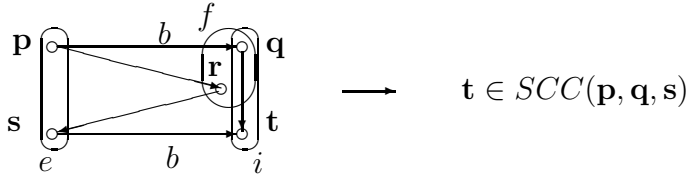
  \begin{picture}(180,85)

\put(45,29){\circle{4}}
      \put(28,30){${\bf s}$}

 \put(104,51){$\bf r$}
\put(88,16){$b$}
\put(85,63){$b$}
 \put(108,45){\vector(-4,-1){61}}
\put(109,46){\circle{4}}
\put(47,62){\vector(1,0){68}}
\put(47,29){\vector(1,0){68}}

\put(117,62){\circle{4}}
  \put(126,60){$\bf q$}
\put(117,60){\vector(0,-1){30}}
 \put(47,62){\vector(4,-1){62}}
\put(45,62){\circle{4}} \put(30,61){$\bf p$}
\put(117,29){\circle{4}}\put(126,30){$\bf t$}

\put(46,46){\oval(10,46)}
\put(112,54){\oval(20,30)}
 \put(40,15){$e$}
 \put(100,70){$f$}
\put(118,46){\oval(10,46)}
\put(121,14){$i$}

 \put(155,45){\vector(1,0){20}}
  \put(200,42){${\bf t} \in SCC({\bf p, q, s})$}
 \end{picture}\\
By Lemma \ref{6}, $SCC({\bf p, q, s})$ is well-defined for transition graphs of
locally threshold testable automata.
\begin{lem} \label{7}
  Let $({\bf p, r}_1)$ and $({\bf p, r}_2)$ be cycle states
of the graph $\Gamma^2$.
Suppose that ${\bf r}_1 \sim {\bf r}_2$,
 ${\bf q}  \succeq {\bf t}_i$,
${\bf p}  \succeq {\bf r} \succeq {\bf r}_i$ ($i=1,2$) for some ${\bf r}$
such that  $({\bf q, r})$ is a cycle state.
Then ${\bf t}_1 \sim {\bf t}_2$
and $SCC({\bf p,q, r}_1) = SCC({\bf p,q, r}_2)$.
   \end{lem}
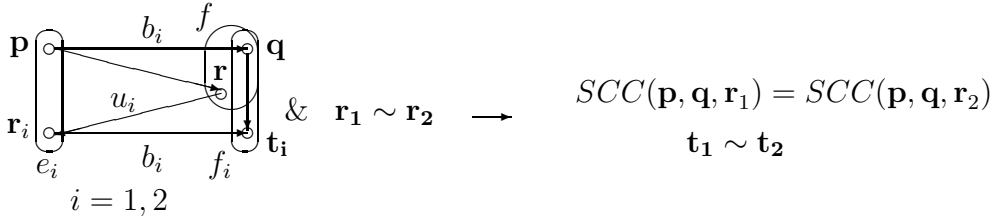
\begin{picture}(100,70)
\put(16,30){\circle{4}}
      \put(0,30){${\bf r}_i$}
   \put(98,23){$\bf t_i$}

 \put(18,30){\vector(1,0){73}}
\put(91,30){\circle{4}}
  \put(51,17){$b_i$}

\put(91,62){\circle{4}}
 \put(98,60){$\bf q$}
\put(91,60){\vector(0,-1){29}}

 \put(18,62){\vector(1,0){73}}
\put(16,62){\circle{4}}
 \put(1,61){$\bf p$}
  \put(51,66){$b_i$}

\put(16,46){\oval(10,46)}
\put(90,46){\oval(10,46)}
\put(85,55){\oval(20,32)}

 \put(11,15){$e_i$}
 \put(76,15){$f_i$}
\put(71,71){$f$}

\put(81,45){\circle{4}}
\put(78,50){${\bf r}$}

\put(81,45){\vector(-4,-1){63}}
\put(18,62){\vector(4,-1){63}}

\put(39,40){$u_i$}
 \put(24,1){$i=1,2 $}

 \put(105,35){$\&$}
  \put(124,35){$\bf r_1 \sim r_2$}

\put(176,36){\vector(1,0){15}}
  \put(215,42){$SCC({\bf p,q, r}_1) = SCC({\bf p,q, r}_2)$}
\put(257,23){$\bf t_1 \sim t_2$}

 \end{picture}\\
  Proof.
 One has $({\bf p, r}_1)e_1=({\bf p, r}_1)$,
$({\bf p, r}_2)e_2=({\bf p, r}_2)$, $({\bf q, r})f=({\bf q, r})$
     for some idempotents $e_1$, $e_2$, $f \in S$,
${\bf p}\succeq {\bf r}$
 and   $({\bf p,r}_1)b_1=({\bf q, t}_1)$,
   $({\bf p,r}_2)b_2=({\bf q, t}_2)$,
   ${\bf r}u_i={\bf r}_i$
  for some elements $u_i$, $b_1$, $b_2 \in S$.
\\
   From $({\bf p,r}_1)e_2=({\bf p, r}_1e_2)$,
by Lemma \ref{4},
 it follows that ${\bf r}_1 \sim {\bf r}_1e_2$.
Notice that ${\bf r}_2e_2 = {\bf r}_2 \sim {\bf r}_1$,
whence ${\bf r}_2 \sim {\bf r}_1e_2$.
Therefore, by Lemma \ref{2}, ${\bf r}_2 = {\bf r}_1e_2$.
Furthermore, $({\bf p,r}_1)e_2b_2=({\bf q, r}_1e_2b_2)=
({\bf q, r}_2b_2)= ({\bf q, t}_2)$.
Thus $({\bf p,r}_1) \succeq ({\bf q, t}_2)$.
Then $({\bf p,r}_1) \succeq ({\bf q, t}_1)$
in view of $({\bf p,r}_1)b_1=({\bf q, t}_1)$.
 Now by Lemma \ref{6}, the states ${\bf t}_1, {\bf t}_2$
 belong to $SCC({\bf p,q, r}_1)$ and
 ${\bf t}_1 \sim {\bf t}_2$.
 Let us notice that
the state ${\bf t}_2$ belongs to
$SCC({\bf p,q, r}_2)$ too. Hence by Lemma \ref{6},
$SCC({\bf p,q, r}_1) = SCC({\bf p,q, r}_2)$.
\begin{lem} \label{12}
  If \\
${\bf p}\succeq \bf q$, ${\bf p}\succeq \bf r$ and ${\bf p}e={\bf p}$
for an idempotent $e \in S$,\\
the state $({\bf q, r}$) is a cycle state of
the graph $\Gamma^2$,\\
there exists a state ${\bf r}_1$ such that
(${\bf p, r}_1)$ is a cycle state and ${\bf r} \succeq {\bf r}_1$,\\
 then $SCC({\bf p,q,r}_1e)=SCC({\bf p,q,r}_1)$.
   \end{lem}
\begin{picture}(170,41)

\put(25,10){\circle{4}}
      \put(7,8){${\bf r}_1$}
 \put(95,13){$\bf r$}
 \put(89,24){\vector(-4,-1){62}}

\put(91,25){\circle{4}}
\put(27,42){\vector(1,0){63}}

\put(91,42){\circle{4}}  \put(99,40){$\bf q$}
 \put(27,42){\vector(4,-1){63}}
\put(8,41){$\bf p$}
\put(25,42){\circle{4}}

\put(25,26){\oval(14,46)}
\put(91,34){\oval(12,30)}

\put(107,22){$\&$ ${\bf p}e={\bf p}$}
 \put(168,24){\vector(1,0){12}}
  \put(184,22){$SCC({\bf p, q, r}_1) =  SCC({\bf p, q, r}_1e)$}
 \end{picture}\\
Proof.
One has ${\bf p}\succeq {\bf r} \succeq {\bf r}_1$.
Then (${\bf p, r}_1) \succeq ({\bf p, r}_1)e = ({\bf p, r}_1e)$ and both these
states are cycle states. Therefore, by Lemma \ref{4},
 ${\bf r}_1e \sim {\bf r}_1$.
 Lemma \ref{7} for ${\bf r}_2 = {\bf r}_1e$
 implies now $SCC({\bf p,q,r}_1e)=SCC({\bf p,q, r}_1)$.
 \begin{thm} $\label{14}$
$DFA$ {\bf A} with state transition complete graph $\Gamma$
(or completed by a sink state) is locally threshold testable iff
\\
\begin{itemize}
\item 1) for every cycle state ($\bf p, q$) of $\Gamma^2$, ${\bf p}
\sim \bf q$ implies ${\bf p} = \bf q$,
\\
\item 2) for every states ${\bf p, q, t, s}$ of
 $\Gamma$ such that
\begin{itemize}
\item (${\bf p, s})$ is a cycle state,
\item $({\bf p,s}) \succeq ({\bf q, t})$,
\item ${\bf p} \succeq {\bf r} \succeq {\bf s}$ and
(${\bf r, t}$) is a cycle state for some ${\bf r}$,
 \end{itemize}
it holds ${\bf q} \succeq {\bf t}$. (see figure to Lemma \ref{5})
\\
\item 3) for every states ${\bf p, q, r}$, $SCC({\bf p, q, r})$
is well defined,
\\
 \item 4) for every four states ${\bf p, q, r, q}_1$ such that
\begin{itemize}
\item (${\bf p, q}_1)$ and $({\bf q, r}$) are cycle states of
the graph $\Gamma^2$,
\item ${\bf p}\succeq \bf q$ and ${\bf p}\succeq \bf r$,
\item for some state ${\bf r}_1$ such that
(${\bf p, r}_1)$ is a cycle state and
$({\bf q,r})\succeq ({\bf q}_1,{\bf r}_1)$,
\end{itemize}
 it holds $SCC({\bf p,q,r}_1)=SCC({\bf p,r,q}_1)$.
\end{itemize}
\end {thm}
\begin{picture}(170,79)
\put(25,30){\circle{4}}
      \put(7,28){${\bf r}_1$}

 \put(102,44){$\bf r$}

 \put(89,44){\vector(-4,-1){62}}
 \put(90,60){\vector(-4,-1){54}}
\put(52,44){$u$}
\put(52,28){$u$}
\put(90,45){\circle{4}}
\put(27,62){\vector(1,0){63}}

\put(92,62){\circle{4}}  \put(100,60){$\bf q$}
 \put(27,62){\vector(4,-1){63}}
\put(25,62){\circle{4}} \put(8,61){$\bf p$}
\put(36,45){\circle{4}}\put(19,45){${\bf  q}_1$}

\put(25,46){\oval(14,46)}
\put(29,53){\oval(24,30)}
 \put(20,14){$i$}
 \put(35,69){$e$}
\put(91,54){\oval(12,30)}
\put(93,28){$f$}

 \put(122,45){\vector(1,0){15}}
  \put(155,42){$SCC({\bf p, r, q}_1) =  SCC({\bf p, q, r}_1)$}
 \end{picture}\\
 Proof. Let $\bf A$ be a locally threshold testable DFA.
  Condition 1 follows in this case from Lemma \ref{2}.
Condition 2 follows from Lemma \ref{5}.
 Condition 3 follows from Lemma \ref{6}.
\\
Let us check the last condition.
For some idempotent $e$, it holds (${\bf p, q}_1)e = ({\bf p, q}_1)$.
By Lemma \ref{12},
 $SCC({\bf p,q,r}_1e)=SCC({\bf p,q, r}_1)$.
Therefore let us compare $SCC({\bf p,q,r}_1e)$
and $SCC({\bf p,r,q}_1)$.

\begin{picture}(200,84)

\put(25,10){\circle{4}} \put(30,77){$e$}

 \put(27,9){\vector(1,0){63}}
    \put(3,-2){${\bf r_1}e$}

\put(90,10){\circle{4}}
  \put(97,8){${\bf t}_1 \in SCC({\bf p, q, r}_1e)$}
 \put(73,76){$f_1$}
  \put(58,12){b}

\put(25,30){\circle{4}}
 \put(27,30){\vector(1,0){74}}
\put(101,30){\circle{4}}
  \put(105,31){${\bf t} \in SCC({\bf p, r, q}_1)$}
 \put(114,77){$f$}

  \put(43,31){a}
  \put(6,30){$\bf q_1$}
 \put(88,50){\vector(-3,-2){60}}  \put(97,50){$\bf r$}
 \put(88,49){\vector(-3,-2){54}}
\put(90,50){\circle{4}}
 \put(34,20){u}

\put(90,70){\circle{4}}  \put(97,68){$\bf q$}

 \put(27,71){\vector(1,0){62}}
\put(25,70){\circle{4}} \put(10,71){$\bf p$}
  \put(58,74){b}
 \put(88,70){\vector(-3,-2){60}}
 \put(88,69){\vector(-3,-2){54}}
 \put(48,49){u}
 \put(27,70){\vector(3,-1){61}}
  \put(54,62){a}

\put(25,40){\oval(10,70)}
\put(89,40){\oval(10,70)}
\put(99,53){\oval(30,56)}

 \put(155,50){\vector(1,0){20}}
  \put(183,46){$SCC({\bf p, q, r}_1e)=SCC({\bf p, r, q}_1)$}

 \end{picture}

One has ${\bf t}_1f = {\bf r}_1ebf = {\bf p}eafuebf$.
Then by (\ref{e1}) ${\bf p}eafuebf = {\bf p}ebfueaf = {\bf q}ueaf =
{\bf q}_1af = \bf t$. So ${\bf t}_1 \succeq \bf t$. Analogously,
${\bf t} \succeq {\bf t}_1$. Therefore ${\bf t}_1 \sim \bf t$, whence
$SCC({\bf p, r, q}_1) = SCC({\bf p, q, r}_1e)$.
Consequently, $SCC({\bf p, q, r}_1)=SCC({\bf p,r, q}_1)$.
\\
 Conversely, suppose that all four conditions of the theorem hold.
Our aim is to prove the local threshold
testability of DFA. For this aim, let us consider an arbitrary state
$\bf v$, arbitrary elements $a, u, b$ and idempotents $e, f$ from
the syntactic semigroup $S$ of the automaton.
We must to prove that ${\bf v}eafuebf = {\bf v}ebfueaf$ (Theorem \ref{1}).
\\
Let us denote ${\bf p} = {\bf v}e$, ${\bf q} = {\bf v}ebf$,
${\bf q}_1 = {\bf v}ebfue$, ${\bf t} = {\bf v}ebfueaf$,
${\bf r} = {\bf v}eaf$, ${\bf r}_1 = {\bf v}eafue$,
${\bf t}_1 = {\bf v}eafuebf$.
\\
We have $({\bf p, r}_1) \succeq ({\bf q, t}_1)$, the states
$({\bf p, r}_1)$, $({\bf q, t}_1)$ and $({\bf r, t}_1)$ are cycle states,
${\bf p} \succeq {\bf r} \succeq {\bf r}_1$.
Therefore by condition 2, for ${\bf r}_1 = \bf s$, ${\bf q} \succeq {\bf t}_1$.
Now ${\bf t}_1 \in SCC({\bf p, q, r}_1)$. Analogously
${\bf t} \in SCC({\bf p,r, q}_1)$.
 The state
 (${\bf p, q}_1)$ is a cycle state and $({\bf q,r})ue = ({\bf q}_1,{\bf r}_1)$.
 Hence condition 4 implies
$SCC({\bf p, q, r}_1)=SCC({\bf p,r, q}_1)$.
These sets are well-defined, whence by condition 3, ${\bf t}_1 \sim {\bf t}$.
Both these states have common right unit $f$.
Consequently, $({\bf t, t}_1)$ is a cycle state. Now by condition 1,
 ${\bf t}_1 = {\bf t}$. Thus ${\bf v}eafuebf = {\bf v}ebfueaf$
is true for an arbitrary state $\bf v$
 and the identity $eafuebf = ebfueaf$
of local threshold testability holds.
\\
It remains now to prove the aperiodicity of $S$.
 Let $\bf p$ be an arbitrary state and let
$s$ be an arbitrary element of $S$. The semigroup $S$
 is finite, whence for some integers $k$ and $m$, it
holds $s^k=s^{k+m}$. Let us consider the states ${\bf p}s^k$ and
${\bf p}s^{k+1}$. We have ${\bf p}s^k \succeq {\bf p}s^{k+1}$ and, in view
$s^k=s^{k+m}= s^{k+1}s^{m-1}$, it holds ${\bf p}s^{k+1} \succeq {\bf
p}s^k$. Thus ${\bf p}s^{k+1} \sim {\bf p}s^k$. Some power of $s$ is
an idempotent and a right unit for both these states. Then by
condition 1, ${\bf p}s^k = {\bf p}s^{k+1}$. Therefore
$S$ is aperiodic,
and thus the automaton is locally threshold testable.
\begin{lem} \label{15}
 Let $P({\bf q, r})$ be a non-empty set of cycle states
 of a locally threshold testable $DFA$
such that
${\bf p}  \succeq {\bf q}$ and ${\bf p}\succeq \bf r$
for a cycle state $({\bf q, r})$.
\\
By ${\bf r}_2 \rho_r {\bf r}_1$ we denote the case that
for a pair of cycle states $({\bf p},{\bf r}_1)$ and $({\bf p},{\bf r}_2)$,
it holds $({\bf q, r}) \succeq ({\bf q}_1,{\bf r}_1)$ and
$({\bf q, r}) \succeq ({\bf q}_1,{\bf r}_2)$.
\\
Then ${\bf r}_1 \rho_r {\bf r}_2$ implies
 $SCC({\bf p,q, r}_1) = SCC({\bf p,q, r}_2)$
for any ${\bf p} \in P({\bf q, r})$.
   \end{lem}

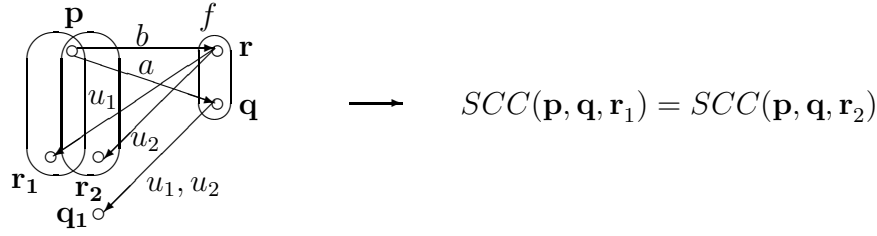
\begin{picture}(200,86)

\put(2,18){$\bf r_1$}
    \put(26,15){$\bf r_2$}
\put(17,30){\circle{4}}
\put(35,30){\circle{4}}
\put(73,80){$f$}

  \put(35,8){\circle{4}}

  \put(47,34){$u_2$}
\put(53,17){$u_1, u_2$}
\put(31,50){$u_1$}
\put(50,61){$a$}
\put(49,72){$b$}

 \put(19,5){$\bf q_1$}

\put(80,50){\circle{4}}
\put(88,46){$\bf q$}
\put(80,70){\circle{4}}
  \put(88,68){$\bf r$}
 \put(27,71){\vector(1,0){52}}

\put(25,70){\circle{4}}
 \put(22,81){$\bf p$}

 \put(79,71){\vector(-3,-2){61}}
 \put(79,71){\vector(-1,-1){42}}
\put(79,51){\vector(-1,-1){42}}

 \put(26,68){\vector(3,-1){53}}

\put(19,50){\oval(22,54)}
\put(32,50){\oval(22,54)}
\put(79,60){\oval(12,32)}

 \put(130,50){\vector(1,0){20}}
  \put(172,46){$SCC({\bf p, q, r}_1)=SCC({\bf p, q, r}_2)$}

 \end{picture}
\\
Proof. One has $({\bf q, r})f = ({\bf q, r})$,
$({\bf q},{\bf r})u_1 = ({\bf q}_1,{\bf r}_1)$,
 $({\bf q},{\bf r})u_2 = ({\bf q}_1,{\bf r}_2)$,
${\bf p}a = {\bf q}$, ${\bf p}b = {\bf r}$,
${\bf p}e=\bf p$ for some idempotents $e$, $f$ and elements $u_i$,
$a$, $b$ from $S$.
  So ${\bf q}_1 = {\bf p}eafu_2 = {\bf p}eafu_1$, ${\bf p}ebfu_1 = {\bf r}_1$,
${\bf p}ebfu_2 = {\bf r}_2$.
For the state ${\bf r}_1eaf$ from $SCC({\bf p,q, r}_1e)$, it holds
${\bf r}_1eaf = {\bf p}ebfu_1eaf = {\bf p}eafu_1ebf =
 {\bf p}eafu_2ebf = {\bf p}ebfu_2eaf = {\bf r}_2eaf \in SCC({\bf p,q, r}_2e)$.
So $SCC({\bf p,q, r}_1e) = SCC({\bf p,q, r}_2e)$.
Thus ${\bf r}_2 \rho_r {\bf r}_1$ implies
$SCC({\bf p,q, r}_1e) = SCC({\bf p,q, r}_2e)$.
By Lemma \ref{12}, $SCC({\bf p,q, r}_ie) = SCC({\bf p,q, r}_i)$,
whence $SCC({\bf p,q, r}_1) = SCC({\bf p,q, r}_2)$.

\begin{cor} \label{l6}
 Let $P({\bf q, r})$ be a non-empty set of cycle states $\bf p$
of a locally threshold testable $DFA$ such that
${\bf p}  \succeq {\bf q}$ and ${\bf p}\succeq \bf r$
for cycle state $({\bf q, r})$.
\\
Then non-empty $SCC({\bf p,q, r}_1)$ does not depend on ${\bf r}_1$
for any ${\bf p} \in P({\bf q, r})$.
   \end{cor}
     \section{An algorithm for local threshold
testability}
   A linear depth-first search algorithm which finds
 all $SCC$ (see \cite {Ta}) will be used.
      The algorithm is based on Theorem \ref{14}
for a complete transition graph $\Gamma$
 (or $\Gamma$ which is completed by sink state).
The measures of complexity of the transition graph
$\Gamma$ are here $|\Gamma|$ (state complexity),
the sum of the numbers of the states and the transitions
 $\it k$ and the size of the alphabet $\it g$ of the labels.
 Let us notice that $|\Gamma|(g+1) \ge k$.
\\
    Let us find all $SCC$ of the graphs $\Gamma$
and $\Gamma^2$ and all their cycle states.
   Furthermore we recognize the reachability on the graph $\Gamma$
 and form the table of reachability for all pairs
of states. The step uses $O(|\Gamma|^2g)$ time and space.
\\
   {\it The first condition of Theorem \ref{14}.}
  For every cycle state ($\bf p, q$) (${\bf p} \neq \bf q$)
from $\Gamma^2$, let us check
  the condition ${\bf p} \sim \bf q$.  A negative answer for
any considered
 cycle state ($\bf p, q$) implies the validity of the condition. In
the opposite case, the automaton is not locally threshold testable.
   The time of the step is $O(|\Gamma|^2)$.
\\
{\it The second condition of Theorem \ref{14}.}
For every cycle state $({\bf p,s})$, we form the set $T$
of states ${\bf t} \in \Gamma$ such that ${\bf s} \succeq {\bf t}$
 and for some state $\bf r$ holds: (${\bf r, t}$)
is a cycle state and ${\bf p} \succeq {\bf r} \succeq {\bf s}$.
If there exists a state ${\bf q}$ such that
$({\bf p, s}) \succeq ({\bf q, t}$) for ${\bf t} \in T$ and
${\bf q} \not\succeq {\bf t}$, then
the automaton is not threshold locally testable.
It is a step of worst case asymptotic cost $O(|\Gamma|^4g)$
with space complexity $O(|\Gamma|^3)$.
\\
{\it The condition 3 of Theorem \ref{14}.}
 For every three states ${\bf p, q, s}$ of the automaton
such that $({\bf p,s})$ is a cycle state,
${\bf p}  \succeq {\bf s}$
and ${\bf p}\succeq {\bf q}$,
let us find a state $\bf r$ such that
 ${\bf p} \succeq {\bf r} \succeq {\bf s}$
and then let us find $SCC({\bf p, q, s})$.
 In the case that this set is not well-defined
(for ${\bf t}_1$, ${\bf t}_2$ from $SCC({\bf p, q, s})$
${\bf t}_1 \not\sim {\bf t}_2$), the
automaton is not threshold locally testable (Lemma \ref{6}).
   The time required for this step in the worst case
is $O(|\Gamma|^4g)$. The space complexity is $O(|\Gamma|^3)$.
\\
Before checking condition 4, let us check the assertion of Lemma \ref{15}.
 For every cycle state $({\bf q, r})$ of the graph $\Gamma^2$, let us form the
set $P({\bf q, r})$ of cycle states $\bf p$ such that ${\bf p}
\succeq {\bf q}$ and ${\bf p}\succeq \bf r$. We continue for non-empty set
$P({\bf q, r})$. For every state ${\bf q}_1$, let us
form the set $R(q_1)$ of states ${\bf r}_1$ such that $({\bf q, r})
\succeq ({\bf q}_1,{\bf  r}_1)$ and the state $({\bf q}_1,{\bf  r}_1)$
 is a cycle state.
 Let us consider two states ${\bf r}_1$, ${\bf r}_2$ from the set $R(q_1)$
such that the states
  $({\bf p, r}_1)$ and $({\bf p},{\bf  r}_2)$ are cycle states
for some $\bf p$ from $P({\bf q, r})$ .
If $SCC({\bf p, q,r}_1) \ne SCC({\bf p, q, r}_2)$, then
the automaton is not locally threshold testable.
\\
{\it The condition 4 of Theorem \ref{14}.}
For every cycle state $({\bf q, r})$ of $\Gamma^2$, let us form the
set $P({\bf q, r})$ of cycle states $\bf p$ such that ${\bf p}
\succeq {\bf q}$ and ${\bf p}\succeq \bf r$. We continue for non-empty set
$P({\bf q, r})$. By Corollary \ref{l6},
$SCC({\bf p, q, r}_1)$ for given $\bf r$ depends
only on the states $\bf p, q$, and $SCC({\bf p, r, q}_1)$
for given $\bf q$ depends only on $\bf p, r$.
If $SCC({\bf p, q, r}_1)$ and $SCC({\bf p, r, q}_1)$ exist and
 are not equal, then the automaton is not locally threshold
testable according to condition 4.
The time required for these last two steps in the worst case
is $O(|\Gamma|^4g)$ with $O(|\Gamma|^3)$ space.
\\
  A positive answer for all the cases implies
the local threshold testability of the automaton.
The time complexity of the algorithm
is no more than $O(|\Gamma|^4g)$. The space complexity
is $max(O(|\Gamma|^2g),O(|\Gamma|^3))$.
In more conventional formulation, we have $O(k^4)$ time and
$O(k^3)$ space.
\section{Conclusion. The package TESTAS}
The considered algorithm is now implemented as a part of the $C/C
^{++}$ package TESTAS replacing the old version of the algorithm
and reducing the time of execution. The program worked essentially
faster in many cases we have studied because of the structure of the
algorithm. A part of branches of the algorithm have only
$O(|\Gamma|^2g)$ or $O(|\Gamma|^3g)$ time and space complexity.\\
The maximal size of the considered graphs on an ordinary PC
was about several hundreds states with an alphabet of several dozen
letters. The program in such case used memory on hard disc
 and works slower.
 \\
The package realizes, besides the considered algorithm for local
threshold testability, a set of algorithms for checking local
testability, left local testability, right local testability,
 piecewise testability and some other programs.
The package checks also the synchronizeability of the automaton and
finds synchronizing words.
 The programs of the package TESTAS analyze:
\\
 1) an automaton of the language presented by oriented labelled
graph. The automaton is given by the matrix:
\\
                      \centerline{ states X labels}
\\
  The non-empty (i,j) cell contains the state
 from the end of the transition with label
 from the j-th column and beginning in the i-th state.
\\
 2) An automaton of the language presented  by its syntactic semigroup.
The semigroup is presented by the matrix (Cayley graph)
\\
\centerline{ elements X generators}
\\
where the i-th row of the matrix is a list of products of
the i-th element on all generators.
The set of generators is not necessarily minimal, therefore
the multiplication table of the semigroup (Cayley table) is acceptable, too.
\\
Some auxiliary programs of the package find direct products of the
objects and build the syntactic semigroup of the automaton on the base
of the transition graph.
\\
 The space complexity of the algorithms which consider the
transition graph of an automaton is not less than $|\Gamma|g$
 because of the structure of the input.
 The graph programs use usually a table
of reachability defined on the states of the graph.
The table of reachability is a square table
and so we have $|\Gamma|^2$ space complexity.
\\
 The number of the states of $\Gamma^n$ is $|\Gamma|^n$,
the alphabet is the same as in $\Gamma$.
So the sum of the numbers of the states
and the transitions of the graph $\Gamma^n$
 is not greater than $(g+1)|\Gamma|^n$.
Some algorithms of the package use the powers $\Gamma^2$,
and $\Gamma^3$. So the space complexity of the algorithms
reaches in these cases $|\Gamma|^2\it g$ or $|\Gamma|^3\it g$.
\\
An algorithm for the local testability problem
 for the transition graph (\cite {TC})
 of $O(k^2)$ (or  $O((|\Gamma|^2g)$ ) time and
space is implemented in the package.
An algorithm of $O(|\Gamma|^2g)$ time
and of $O(|\Gamma|^2g)$ space is used for
finding the bounds on the order of local testability for a given
transition graph of the automaton \cite {Tp}.
 An algorithm of worst case $O(|\Gamma|^3g)$ time complexity
and of $O(|\Gamma|^2g)$ space complexity checked the $2$-testability \cite {Tp}.
The $1$-testability is verified using an algorithm \cite {K94} of
 order $O(|\Gamma|g^2)$.
For checking the $n$-testability \cite {Tp}, we use
 an algorithm of worst case asymptotic cost $O(|\Gamma|^3g^{n-1})$
 of time complexity with $O(|\Gamma|^2g)$ space.
The time complexity of the last algorithm grows with $n$
 and in this way we obtain a non-polynomial algorithm for finding
the order of local testability. However, $n \le
log_gM$ where $M$ is the maximal size of the integer in the
computer memory.
\\
The time complexity of the algorithm to verify piecewise testability of
DFA is $O(|\Gamma|^2g)$. The space complexity of the algorithm is $O(k)$
\cite {TC}.
\\
The algorithms for right and left local testability
 for the transition graph are essentially distinct,
moreover, the time complexity of the algorithms differs.
The graph algorithm for
the left local testability problem needs in the worst case
 $O(|\Gamma|^3g)$ time and $O(|\Gamma|^3g)$ space and the algorithm
 for the right local testability problem for transition
 graph of the deterministic finite automaton needs
 $O(|\Gamma|^2g)$ time and space \cite {TL}.
\\
The main measure of complexity for semigroup $S$ is the size
of the semigroup $|S|$. We use also the number of generators
(size of alphabet) $\it g$ and the number of idempotents $\it i$.
\\
Algorithms of the package dealing with the transition semigroup
of an automaton use the multiplication table of the semigroup of
$O(|S|^2)$ space.
Other arrays used by the package present subsemigroups or
subsets of the transition semigroup.
So we usually have $O(|S|^2)$ space complexity.
\\
We implement in the package TESTAS
 a polynomial-time algorithms of $O(|S|^2)$ time complexity for the
 local testability problem and for finding the order of local testability
 for a given semigroup \cite{Ts}.\\
The time complexity of the semigroup algorithm
for both left and right local testability is $O(|S|i)$ \cite{TL}.
The time complexity of the semigroup algorithm
for local threshold testability is $O(|S|^3)$.
Piecewise testability is verified
in $O(|S|^2)$ time \cite{TC}.

\section*{Acknowledgments}
I am grateful to the anonymous referees for helpful and detailed comments
that proved very useful in improving the presentation and style of the paper.

 \end{document}